\documentclass[aps,prr,twocolumn,amsmath,amssymb,superscriptaddress,longbibliography]{revtex4-2}
\usepackage[english]{babel}
\usepackage{xcolor}
\usepackage{amssymb} 
\usepackage{dcolumn}
\usepackage{bm}
\usepackage{graphicx}
\usepackage{amsmath}
\usepackage{braket}

\usepackage{graphicx}        

\graphicspath{{pict/}{}}

\usepackage[normalem]{ulem}

\usepackage{dcolumn}
\usepackage{bm}
\usepackage[pdfstartview=FitH, CJKbookmarks=true, bookmarksnumbered=true, bookmarksopen=true, colorlinks=true, pdfborder=001, citecolor=blue, linkcolor=blue, urlcolor=blue, linktocpage=true] {hyperref}

\newcommand{\hN}{\hat{N}}

\newcommand{\hb}{\hat{b}}
\newcommand{\hmm}{\hat{m}}

\begin{document}

\title{Long-Range Three-Body Interaction and Photon-Matter Entanglement with Atom-Molecule Superradiance in an Optical Cavity}

\title{Atom-Molecule Superradiance and Entanglement with Cavity-Mediated Three-Body Interactions}

\author{Yun Chen}
\thanks{These authors contributed equally to this work.}
\affiliation{Guangdong Provincial Key Laboratory of Quantum Metrology and Sensing $\&$ School of Physics and Astronomy, Sun Yat-Sen University (Zhuhai Campus), Zhuhai 519082, China}

\author{Yuqi Wang}
\thanks{These authors contributed equally to this work.}
\affiliation{State Key Laboratory of Low Dimensional Quantum Physics, Department of Physics, Tsinghua University, Beijing 100084, China}

\author{Jingjun You}
\affiliation{Guangdong Provincial Key Laboratory of Quantum Metrology and Sensing $\&$ School of Physics and Astronomy, Sun Yat-Sen University (Zhuhai Campus), Zhuhai 519082, China}

\author{Yingqi Liu}
\affiliation{Guangdong Provincial Key Laboratory of Quantum Metrology and Sensing $\&$ School of Physics and Astronomy, Sun Yat-Sen University (Zhuhai Campus), Zhuhai 519082, China}

\author{Su Yi}
\email{yisu@nbu.edu.cn}
\affiliation{Institute of Fundamental Physics and Quantum Technology $\&$ School of Physics, Ningbo University, Ningbo, 315211, China}
\affiliation{Peng Huanwu Collaborative Center for Research and Education, Beihang University, Beijing 100191, China}

\author{Yuangang Deng}
\email{dengyg3@mail.sysu.edu.cn}
\affiliation{Guangdong Provincial Key Laboratory of Quantum Metrology and Sensing $\&$ School of Physics and Astronomy, Sun Yat-Sen University (Zhuhai Campus), Zhuhai 519082, China}

\date{\today}

\begin{abstract}
Ultracold atoms coupled to optical cavities offer a powerful platform for studying strongly correlated many-body physics. Here, we propose an experimental scheme for creating biatomic molecules via cavity-enhanced photoassociation from an atomic condensate. This setup realizes long-range three-body interactions mediated by tripartite cavity-atom-molecule coupling. Beyond a critical pump strength, a self-organized square lattice phase for molecular condensate emerges, resulting in hybrid atom-molecule superradiance with spontaneous $U(1)$ symmetry breaking. Distinct from previously observed ultracold bosonic (fermionic) atomic superradiance, our findings demonstrate bosonic enhancement characterized by a cubic scaling of steady-state photon number with total atom number. Additionally, strong photon-matter entanglement is shown to effectively characterize superradiant quantum phase transition. Our findings deepen the understanding of quantum superchemistry and exotic many-body nonequilibrium dynamics in cavity-coupled quantum gases.
\end{abstract}

\maketitle
{\em Introduction}.---Engineering novel multibody interactions is a cornerstone in studying correlated many-body quantum phenomena~\cite{RevModPhys.80.885,RevModPhys.82.2313,RevModPhys.85.553}. Ultracold molecules with rich internal structures and tunable permanent electric dipole moments provide a versatile platform for ultracold chemistry~\cite{Chem.Rev.112.4949-5011,Park2023,PRXQuantum.5.020358}, strongly correlated many-body physics\cite{Micheli2006,Covey2016,PRXQuantum.2.017003,Blackmore_2019,Gregory2021,PhysRevLett.98.060403,PhysRevLett.98.060404,PhysRevLett.103.155302,PhysRevLett.98.060404,PhysRevLett.104.125301}, quantum computing~\cite{PhysRevLett.88.067901,Andre2006,Sawant_2020,Cornish2024,Picard2024}, and fundamental physical laws~\cite{Carr_2009,Chin_2009, doi:10.1126/science.adg4084}. Recent advancements in creating ultracold molecules~\cite{doi:10.1126/science.aau7230, PhysRevLett.124.033401, Schindewolf, Duda2023} encompasses various techniques, ranging from direct laser cooling~\cite{PhysRevLett.116.063005, Zeppenfeld2012, PhysRevX.10.021049, Mitra2020} to collisional resonances via magnetoassociation~\cite{doi:10.1126/science.1163861,RevModPhys.82.1225,doi:10.1126/science.aac6400,PhysRevLett.124.253401} or photoassociation (PA)~\cite{RevModPhys.78.483,PhysRevLett.95.063202, doi:10.1126/science.aar7797,PhysRevLett.132.093403}. These techniques have enabled breakthroughs such as controlling hyperfine states~\cite{PhysRevLett.116.225306, PhysRevX.9.011028,PhysRevLett.114.205302,PhysRevLett.116.205303,PRXQuantum.5.020344}, observing dipolar spin-exchange interactions~\cite{Li2023,Christakis2023}, achieving microwave shielding~\cite{PhysRevLett.121.163401, doi:10.1126/science.abg9502,Chen2023} and polar molecular condensates~\cite{bigagli2024observation}. These advances significantly enhance controlling long-range interactions and suppressing inelastic losses to study hitherto unexplored physical phenomena and quantum matter.

Meanwhile, ultracold atoms in optical cavities have offered significant opportunities to study quantum many-body physics with diverse applications~\cite{Ritsch_2021}. Corresponding to the superradiance quantum phase transition (QPT)~\cite{doi:10.1126/science.abd4385,Baumann2010}, a wide range of fundamental quantum phenomena have been studied, including roton-type mode softening~\cite{doi:10.1126/science.1220314}, supersolid phase~\cite{Leonard2017,leonard2017monitoring,PhysRevResearch.5.013002}, and dynamical spin-orbit coupling~\cite{PhysRevLett.121.163601,PhysRevLett.123.160404}. Importantly, mechanism for generating bipartite quantum entanglement has been demonstrated using dipolar interactions or cavity mediated two-body interactions~\cite{PhysRevLett.126.113401, doi:10.1126/science.adf8999, doi:10.1126/science.adf4272,PhysRevLett.132.093402}, with applications in quantum sensing and enhanced matter-wave interferometry. Despite these advances, the superradiance in nonequilibrium dynamics for hybrid atom-molecule system remains unexplored. This regime holds great promise for engineering multibody interactions and tripartite entanglement with intriguing properties.

In this work, we propose the utilization of PA from ultracold atom pairs to create weakly bound electronically diatomic molecule, which are subsequently transfered to stable rovibrational ground state within optical cavity~\cite{PhysRevLett.93.073002, PhysRevLett.95.063202}. We demonstrate that the self-ordered square lattice (SQL) phase for ground-state molecule arises from the atom-molecule superradiance. The SQL phase exhibits an undamped gapless Goldstone mode, a hallmark of spontaneous $U(1)$ symmetry breaking. Unlike earlier explorations that focused on cavity-mediated two-body interactions~\cite{Baumann2010,doi:10.1126/science.1220314}, our approach achieves long-range three-body interactions for matter-wave fields. Particularly intriguing is the bosonic enhancement observed in atom-molecule superradiance, evidenced by a cubic scaling of steady-state photon number ($N_s$) with the total atoms number ($N$), which markedly differs from the experimentally observed ultracold bosonic (fermionic) atomic superradiance with $N_s\sim N^2$ ($N$) due to distinct quantum statistics~\cite{doi:10.1126/science.abd4385,Baumann2010,doi:10.1126/science.1220314,Leonard2017, leonard2017monitoring}. The inherent leakage of cavity superradiance offers a precise benchmark for measuring ultracold molecules, addressing a longstanding challenge in molecular detection. Additionally, the collective enhancement effects give rise to strong cavity-matter entanglement, as quantified by the entropy of entanglement, which fully characterizes the atom-molecule superradiance. Our proposal unveils a novel pathway for realizing controllable three-body interactions, with prospects for exploring fundamental phenomena in strongly correlated physics and emerging photon-matter entanglement in quantum metrology.

\begin{figure}
\includegraphics[width=0.9\columnwidth]{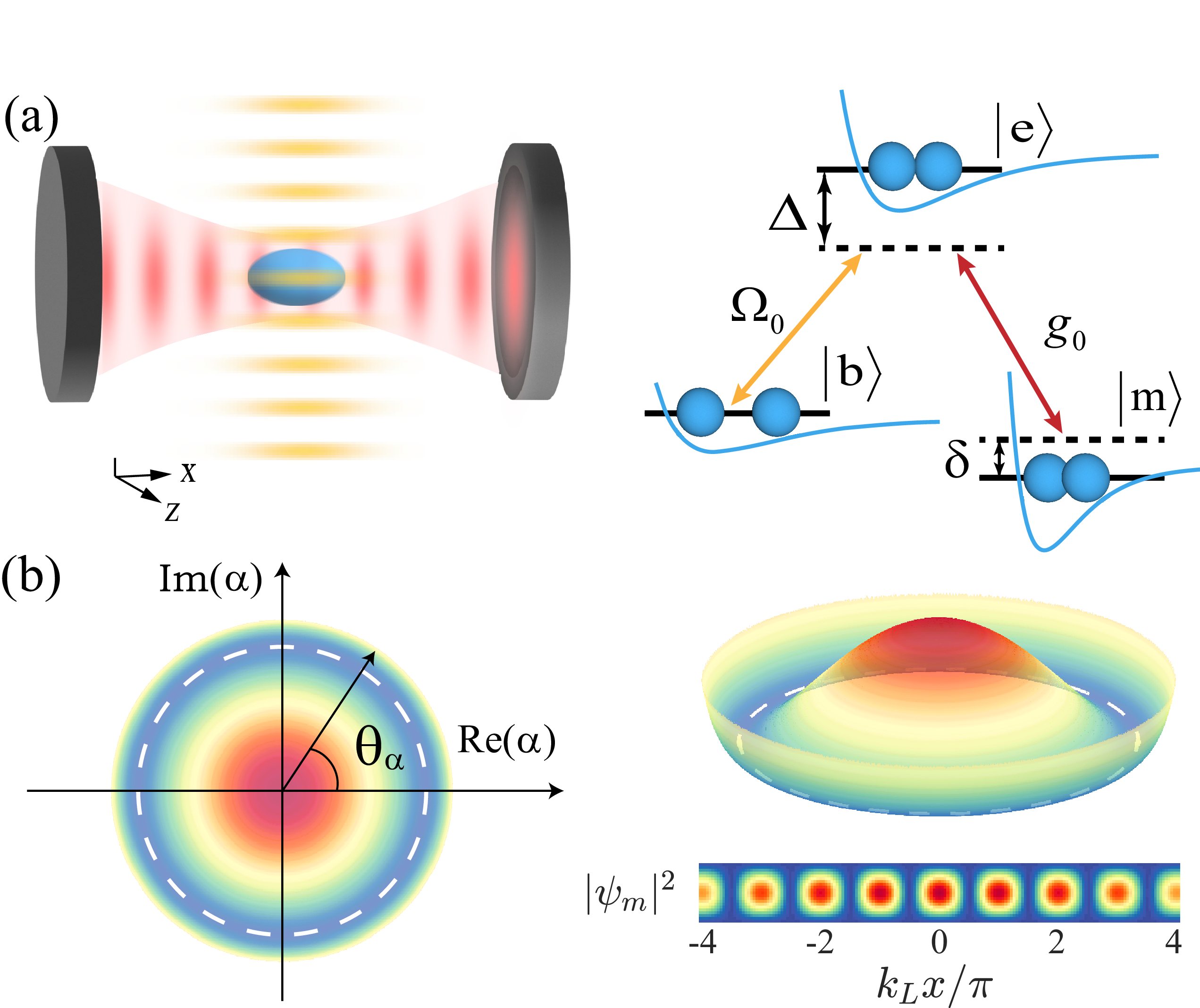} \caption{\label{model}
(a) Scheme for creating atom-molecule superradiance and relevant energy levels for free-bound-bound transitions.  (b) The self-ordered SQL phase exhibiting $U(1)$ symmetry breaking of molecular phases, corresponding to arg($\alpha$)-independent density distribution of molecular wave function.}
\end{figure}

{\em Model}.---We identify an experimental scheme of generating atom-molecule superradiance using a gas of $N$ ultracold $^{133}$Cs atoms inside an optical cavity, as illustrated in Fig.~\ref{model}(a). The cavity decay rate is $\kappa=50 E_L/\hbar$ with $E_L/\hbar=1.8~{\rm kHz}~(2\pi)$ being the single-photon recoil energy. Pairs of atoms are converted into a ground state homonuclear diatomic molecule through cavity-enhanced two-photon PA, involving free-quasi-bound-bound transition. The transition between atomic state $|b\rangle$ and weakly bound molecules $|e\rangle$ is coupled by a transverse standing-wave laser propagating along $y$ axis, with the Rabi coupling $\Omega_0(y)=\Omega_0\cos(k_{L}y)$, where $k_{L}={ 2 \pi}/\lambda$ is the wave vector of the laser field with $\lambda$ being the wavelength. The molecular bound-bound transition $\left|e\right\rangle\leftrightarrow\left|m\right\rangle$ is illuminated by cavity along $x$ axis with photon-matter coupling $g_0$. Without loss of generality, we assume that the cavity and classical PA fields have the equal wave vector. 

In the far detuned PA field, $\left|\Delta\right|\gg\{g_0,\Omega_0\}$, the quasibound molecule state $\left|e\right\rangle$ can be  adiabatically eliminated~\cite{RevModPhys.85.553}. The many-body Hamiltonian for hybrid cavity-atom-molecule system is given by
\begin{align}
\hat {\cal H}_0&=
\hbar\Delta_c\hat{a}^{\dag}\hat{a} + \sum_{\sigma}\int d{\bf{r}}\hat{\psi}_{\sigma}^{\dagger}(\mathbf{r})[\frac{{\bf p}^{2}}{2m_{\sigma}}+V_e+U_{\sigma}]\hat{\psi}_{\sigma}(\mathbf{{r}})\nonumber\\
&+\frac{1}{2}\sum_{\sigma\sigma'}\int d{\bf{r}} (g_{\sigma\sigma'}+\lambda_{\sigma\sigma{'}})\hat{\psi}_{\sigma}^{\dagger}(\mathbf{r})\hat{\psi}_{\sigma'}^{\dagger}(\mathbf{{r}})\hat{\psi}_{\sigma'}(\mathbf{{r}})\hat{\psi}_{\sigma}(\mathbf{{r}})\nonumber\\
&+\hbar\Omega\int d{\mathbf{r}} \cos(k_{L}x)\cos(k_{L}y)\hat{a}\hat{\psi}_{b}^{\dagger2}(\mathbf{r})\hat{\psi}_{m}(\mathbf{r}) 
+{\rm H.c.},\label{manyh} 
\end{align}
where $\hat{a}$ and $\hat{\psi}_{b}$ ($\hat{\psi}_{m}$) are the annihilation operators for the cavity and atomic (molecular) fields with masses satisfying $m_m=2m_b$, $\Omega=-{g_{0}\Omega_{0}}/{\Delta}$ is the atom-molecule conversion strength, $\Delta_{c}$ is the pump-cavity detuning, and $U_{m}=\hbar[\delta+U_{0}\cos^{2}(k_{L}x)\hat{a}^{\dagger}\hat{a}]$ with $U_{0}=-{g_{0}^{2}}/{\Delta}$ being optical Stark shift of cavity and $\delta$ being the two-photon detuning.  $V_e=m_{\sigma}\omega_{\perp}^2(x^2+y^2+\gamma^2z^2)/2$ is the external spin independent trapping potential with $\omega_{\perp}=(2\pi)130\rm{Hz}$ being the radial trap frequency and $\gamma=10$ being the trap aspect ratio. The short-range interaction $g_{{\sigma\sigma}}={4\pi\hbar^{2}a_{{\sigma\sigma}}}/{m_{\sigma}}$ and $g_{\sigma\sigma'}={3\pi\hbar^{2}a_{bm}}/{m_{b}}$  with $a_{\sigma\sigma'}$ being $s$-wave scattering lengths for intraspecies ($\sigma=\sigma'$) and interspecies ($\sigma\neq\sigma'$) matter wave fields. In our simulation, we fix $a_{bb}=a_{bm}=50 a_B$ and $a_{mm}=100 a_B$ with $a_B$ being the Bohr radius. In contrast to Feshbach resonances~\cite{RevModPhys.82.1225}, PA field brings an effective two-body interaction for atoms, $\lambda_{bb}= - 2\hbar ({\Omega_{0}^{2}}/{\Delta})\cos^{2}(k_{L}y)$, which can be used to fine tuning the collisional interaction avoiding the atom losses.

In dispersive limit $|\Delta_c|\gg\{g_0,\Omega_0\}$, cavity reaches a steady state much faster than atomic and molecular motions, corresponding to the intracavity amplitude $\alpha={\Omega\Xi}/({i{\kappa}-\tilde{\Delta}_{c}})$ with $\tilde{\Delta}_{c}=\Delta_{c}+U_{0}\int d\mathbf{{r}}\cos^{2}(k_{L}x)|\psi_{m}|^{2}$. The $\Xi=\int d\mathbf{{r}}\cos(k_{L}x)\cos(k_{L}y)\psi_{b}^{2}\psi_{m}^{*}$ is the order parameter that characterizes self-ordered superradiant QPT and determines the configuration of molecular wave function. Different to cavity-mediated two-body interaction for atom superradiance~\cite{Baumann2010,doi:10.1126/science.1220314}, we realize an effective long-range three-body interaction after integrating out cavity field
\begin{eqnarray}\label{manyint}
\hat {\cal H}_1/\hbar=\frac{\chi}{6}\int d\mathbf{{r}}d\mathbf{{r}^{\prime}}\mathcal{D}(\mathbf{{r}},\mathbf{{r}^{\prime}})\hat{\psi}_{m}^{\dagger}(\mathbf{{r}})\hat{\psi}_{b}^{\dagger 2}(\mathbf{{r}^{\prime}})\hat{\psi}_{m}(\mathbf{{r}^{\prime}})\hat{\psi}_{b}^{2}(\mathbf{{r}}),\nonumber\\
\end{eqnarray}
where $\mathcal{D}(\mathbf{r},\mathbf{r^{\prime}})=\cos(k_{L}x)\cos(k_{L}x^{\prime})\cos(k_{L}y)\cos(k_{L}y^{\prime})$ is the long-range potential and $\chi=-{12\tilde{\Delta}_{c}\Omega^{2}}/({\tilde{\Delta}_{c}^{2}+\kappa^{2}})$ is three-body interactions. Notably, the magnitude of $\chi$ is typically on the order of hundreds Hertz even for steady-state photon number $N_s\sim 1$. This is comparable to the effective three-body interactions observed in optical lattice deeply trapped  ultracold bosonic atoms~\cite{Will2010}.  Unlike intrinsic inelastic loss resonances in three-body collisions of quantum gases~\cite{kraemer2006evidence}, the collective three-body interactions can be significantly enhanced by increasing PA field. This enhancement helps stabilize condensates, preventing collapse due to strong two-body dipolar or attractive contact interactions~\cite{PhysRevA.97.011602}. Our method enables precise control of three-body interaction, which may facilitate study of strongly correlated many-body physics~\cite{Johnson_2009}.

 \begin{figure*}
 \includegraphics[width=1.8\columnwidth]{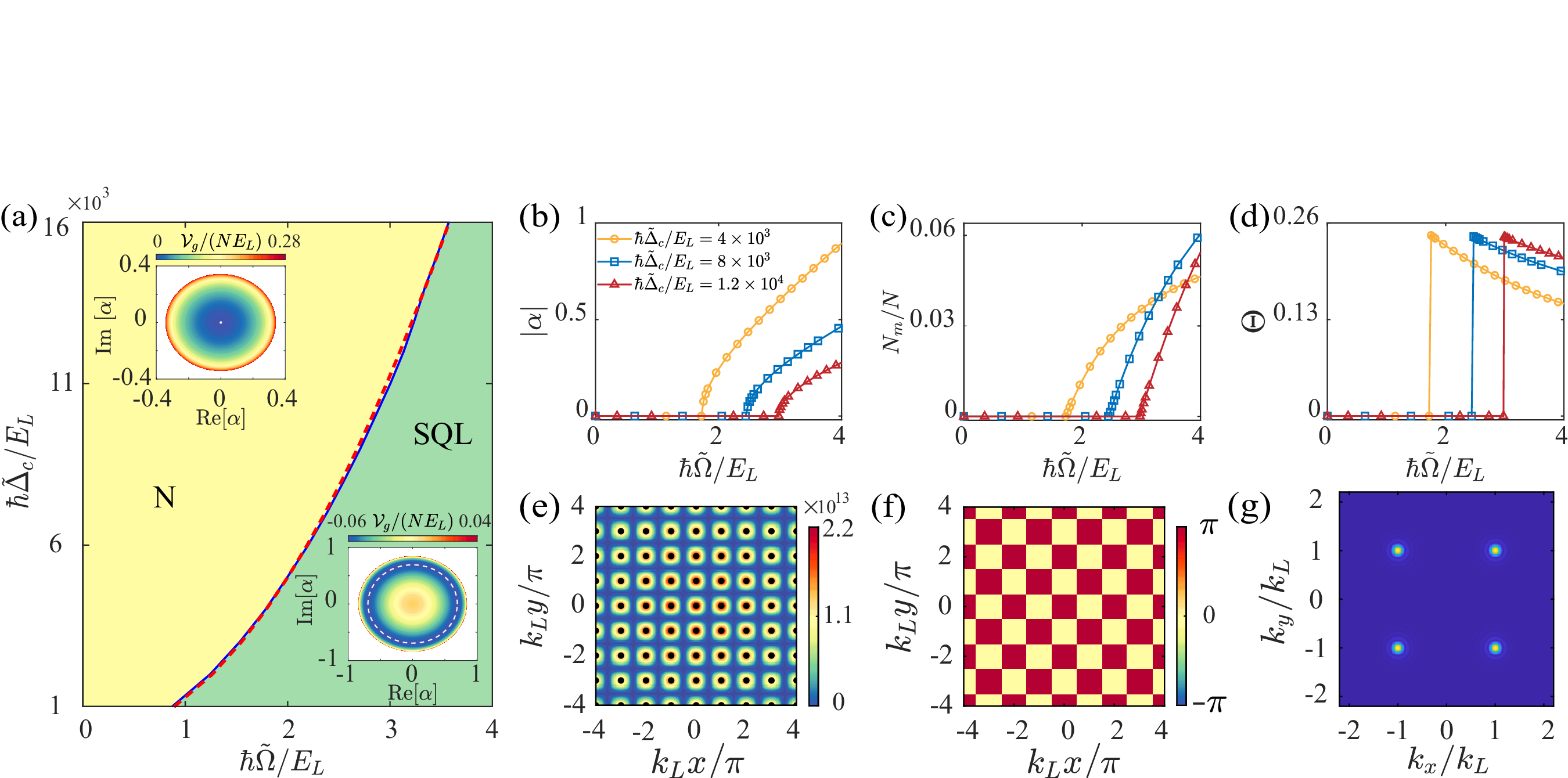}
 \caption{\label{p2_23118}
 (a) Ground state phase diagram for $N=1\times10^4$ and $\hbar\delta/E_L=1 $. The insets show $\alpha$ dependence of ${\cal V}_g$ for N and SQL phases. $\tilde{\Omega}$ dependence of $|\alpha|$ (b),  $N_m$ (c), and $\Theta$ (d) for different $\tilde{\Delta}_c$. The density (e), phase (f), and momentum distribution for SQL phase with $\hbar\tilde{\Omega}/E_L= 2$ and $\hbar\tilde{\Delta}_c/E_L= 4\times 10^3$.}
\end{figure*}

Fundamental insight into the atom-molecule superradiance is understood from the microscopic picture of coherently transfers the atomic motional ground state $|k_{x},k_{y}\rangle=|0,0\rangle$ to molecular excited momentum states $|\pm k_L,\pm k_L\rangle$ via cavity-enhanced two-photon PA process. In the single recoil scattering limit~\cite{Baumann2010,doi:10.1126/science.abd4385}, the many-body Hamiltonian excluding collisional interactions reads
\begin{eqnarray}\label{effh}
\hat {\cal H}_2/\hbar=\tilde{\Delta}_{c}\hat{a}^{\dagger}\hat{a}+\delta^{\prime}m^{\dagger}\hat{m}+\frac{\tilde{\Omega}}{2\sqrt{N}}(\hat{b}^{\dagger}{}^{2}\hat{a}\hat{m}+\rm H.c.),
\end{eqnarray}
where $\hat{b}$ and $\hat{m}$ are annihilation bosonic operators of atom and molecule states with total atom number $\hN=\hb^{\dag}\hb+2\hmm^{\dag}\hmm$, $\delta^{\prime}=\delta+E_{L}/\hbar$, and $\tilde{\Omega}$ is light-matter coupling. This nonlinear tripartite interaction, involving cavity and atom-molecule fields, differs from Dicke model. The Hamiltonian exhibits a $U(1)$ symmetry obeying the commutation relation $[\mathcal{R}_{\theta},\hat{\cal H}_{2}]=0$, where $\mathcal{R}_{\theta}=\exp(i\theta\hat{N}_e)$ and $\hat{N}_e=\hat{m}^{\dag}\hat{m}+\hat{a}^{\dag}\hat{a}+\hat{b}^{\dag}\hat{b}$ is the total excitation number of system. As PA coupling increasing, the system undergoes a QPT from normal $(\alpha=0)$ to superradiant phase $(|\alpha|>0)$, corresponding to $N$-dependent threshold Raman coupling $\tilde{\Omega}_{\rm cr}=2\sqrt{{\delta^{\prime}\tilde{\Delta}_{c}}/{N}}$. Notably, a gapless Goldstone mode of collective excitations is confirmed~\cite{SM}, which is consistent with high ground-state degeneracy resulting from $U(1)$ symmetry breaking.

To proceed further, we derive the effective potential for atom-molecule superradiance
\begin{eqnarray}\label{ev}
{\cal V}({\beta})/\hbar={\frac{\tilde{\Omega}^{2}}{N\tilde{\Delta}_{c}}}(N|\beta|^{4}-|\beta|^{6})+(\delta^{\prime}-{\frac{\tilde{\Omega}^{2}}{4\tilde{\Delta}_{c}}}N)|\beta|^{2},
\end{eqnarray}
 with $\beta=\langle\hat{m}\rangle$. When $\tilde{\Omega}>\tilde{\Omega}_{\rm cr}$ and $\tilde{\Delta}_c>0$, ${\cal V}({\beta})$ transitions from a single minimum at the origin to a sombrero shape potential with a circular valley of degenerate minima [Fig.~\ref{model}(b)], signaling the onset of a second-order QPT. Clearly, the additional term proportional to $|\beta|^{6}$ is emerged for hybrid cavity-atom-molecule system, which may exhibit a first-order QPT for  $\tilde{\Delta}_c<0$.
 
 {\em Ground state structures}.---Next, we explore ground state structures of quantum phases under atom-molecule superradiance. Figure~\ref{p2_23118}(a) shows the phase diagram of atom-molecule cavity system on the $\tilde{\Delta}_{c}$-$\tilde{\Omega}$ parameter plane. The SQL phase originates from atom-molecule superradiance, occurring at small $\tilde{\Delta}_c$ and large $\tilde{\Omega}$. For larger dispersively-shifted cavity detuning, the system remains in N phase when Raman coupling is below the threshold value. The process of self-organization for condensate molecular wave function corresponds to the spontaneous $U(1)$ symmetry breaking from vacuum ($\alpha=0$) to a finite value ($\alpha\neq0$) in steady state. Interestingly, the analytical phase boundary (solid line) between N and SQL is in good agreement with its numerical simulations (dashed line), which demonstrates the complex cavity-atom-molecule superradiance can be fully characterized by tripartite Hamiltonian $\hat {\cal H}_2$. 
 
In Figs.~\ref{p2_23118}(b) and \ref{p2_23118}(c), we plot $\tilde{\Omega}$ dependence of cavity amplitude $|\alpha|$ and molecules number $N_m$ for different values of $\tilde{\Delta}_{c}$. Both $|\alpha|$ and $N_m$ increase rapidly when the Raman coupling exceeds $\tilde{\Omega}_{\rm cr}$. As expected, a large cavity detuning $\tilde{\Delta}_{c}$ respects to a large threshold and relatively small photon number as well. To better characterize superradiant phase, we introduce an order parameter $\Theta=\langle\psi_{m}|\cos(2k_{L}x)\cos(2k_{L}y)|\psi_{m}\rangle/N_m$, which determines periodic density modulation of molecule wave function in ground state. It is clear that $\Theta$ becomes nonzero undergoing the atom-molecule superradiant QPT [Fig.~\ref{p2_23118}(d)]. This indicates the generation of spatial periodicity of crystalline for self-organization SQL phase. The emerged square lattice offers an additional stabilization mechanism for generating long-lived molecules through atom-molecule superradiance~\cite{PhysRevLett.108.080405}.

\begin{figure}
 \includegraphics[width=0.95\columnwidth]{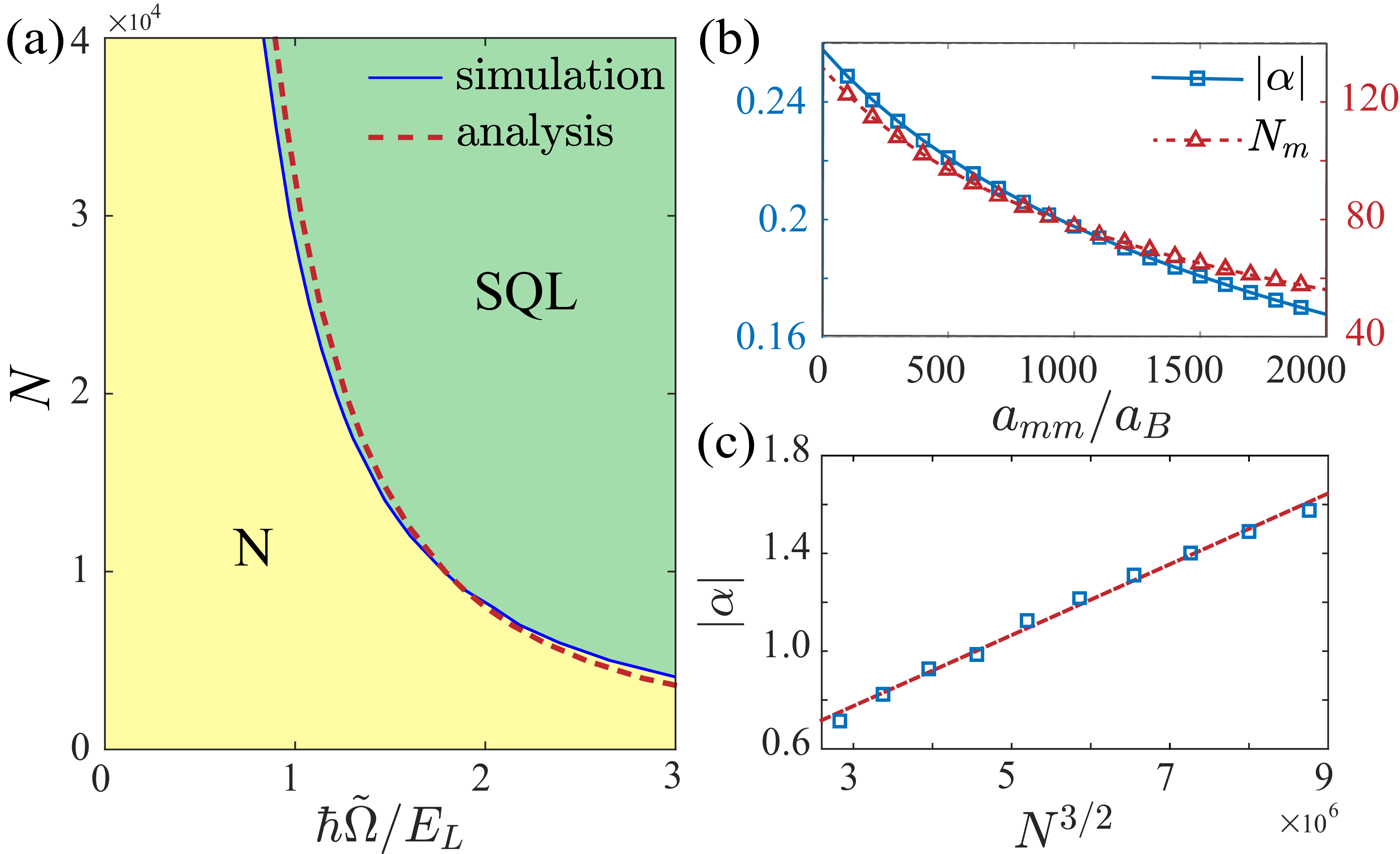}
 \caption{\label{phase2}(a) The phase diagram on $N$-$\tilde{\Omega}$ parameter plane. (b) $a_{mm}$ dependence of $|\alpha|$ (solid line) and $N_m$ (dashed line). (c) $N$ dependence of cavity amplitude. The red dotted line is a fitted straight line proportional to $N^{3/2}$.}
 \end{figure}
 
Figures \ref{p2_23118}(e) and \ref{p2_23118}(g) show the typical density and phase distributions of molecular wave function for SQL phase. To minimize kinetic energy, the atomic condense wave function is structureless along cavity axis and occupies large population in the parameter regime of our numerical simulation. As can be seen, the self-organized SQL phase exhibits a $\lambda/2$-period density modulation along both $x$ and $y$ axes. The max peak density of molecules locate at positions satisfying $\cos^{2}(k_{L}x)\cos^{2}(k_{L}y)=1$. This crystalline structure for density profiles is different from the $\lambda$ periodic ${\cal Z}_2$-broken checkerboard lattice observed in Dicke superradiance~\cite{Baumann2010}. In momentum space, the realized molecular condense at momentums $|\pm \hbar k_L, \pm \hbar k_L\rangle$ [Fig.~\ref{p2_23118}(g)], consistent with the Hamiltonian $\hat {\cal H}_2$ in the single recoil scattering limit. 

Interestingly, the phase of the condensate wave function exhibits a staggered $\lambda$-period phase modulation, as displayed in Fig.~\ref{p2_23118}(f), with a relative phase difference of $\pi$ between neighboring sites. We find that the self-ordered phase profile is directly connected to the cavity phase angle $\rm{arg}(\alpha)$. Importantly, this phase value can change continuously from 0 to $2\pi$, corresponding to atom-molecule superradiance [Fig.~\ref{model}(b)]. The relationship between the phase of the cavity and molecule fields satisfies $\arg(\alpha)+\arg(\psi_m)=0$ ($\pi$) when density is located at positions satisfying $\cos(k_L x)\cos(k_L y)=-1$ ($1$). We  emphasize that SQL phase does not belongs to supersolid phase~\cite{Leonard2017, leonard2017monitoring,PhysRevResearch.5.013002}, as it lacks continuous translational symmetry, despite it possesses a density periodicity of crystalline order and gapless Goldstone mode associated with $U(1)$ symmetry breaking. 

\begin{figure}[ptb]
\includegraphics[width=0.95\columnwidth]{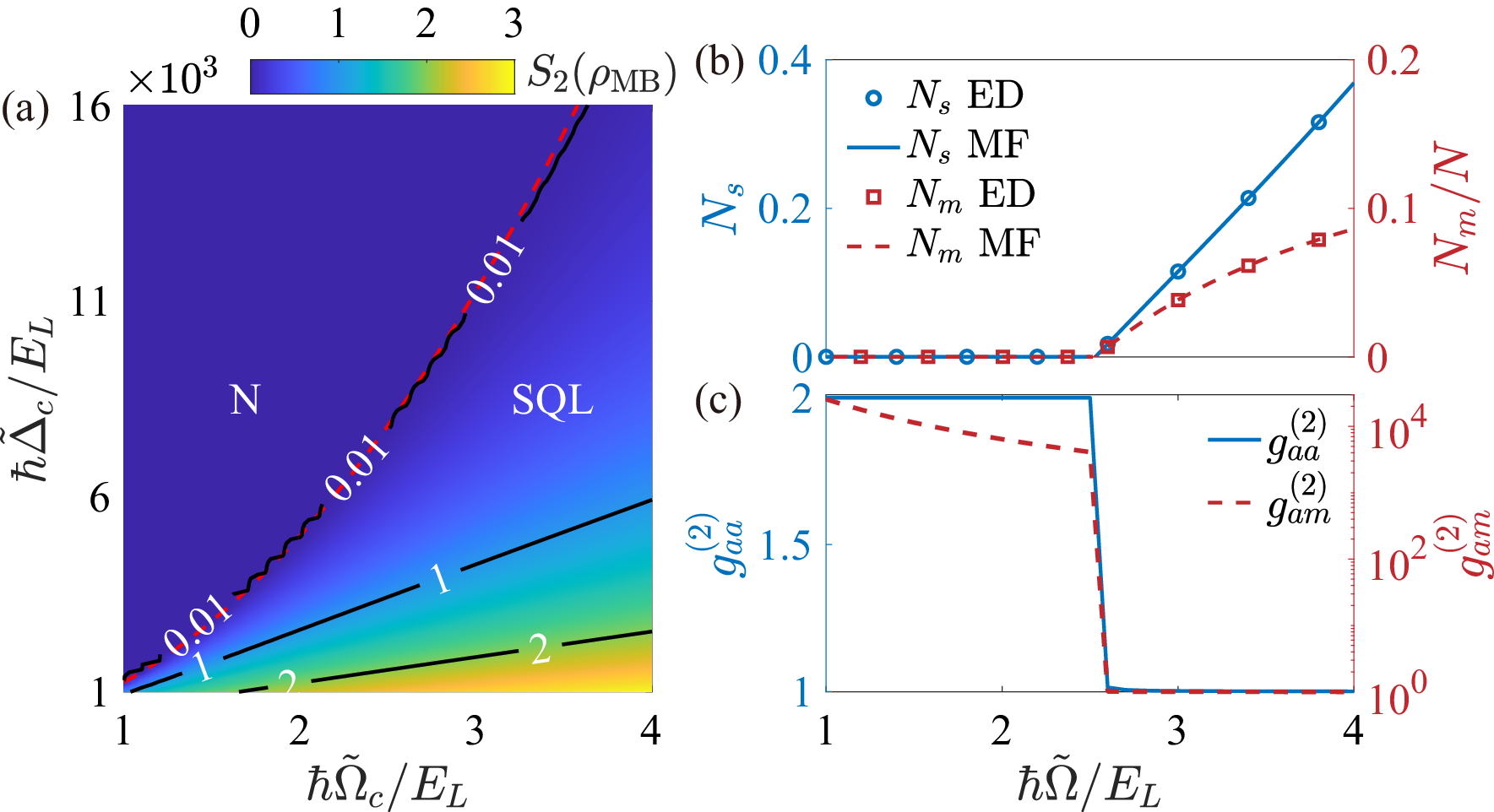} 
\caption{(a) Distributions of entropy $S_2$ on $\tilde \Omega$-$\tilde{\Delta}_c$ parameter plane. (b) $\tilde \Omega$ dependence of $N_s$ (solid line) and $N_m$ (dashed line).  (c) Distribution of  $g^{(2)}_{aa}$ and  $g^{(2)}_{bb}$ as a function of $\tilde \Omega$ for $\hbar\tilde{\Delta}_c/E_L=8\times10^3$.}\label{S2NaNm} 
\end{figure}

{\em Cubic-law scale of superradiance}.---The fundamental insights into the onset of self-ordered superradiance have been derived from cooperative photon emission in collections of ultracold atoms. The effects of Bose (Fermi) statistics manifest in the steady-state photon number scaling as $N^2$ ($N$) with atom number~\cite{Baumann2010,doi:10.1126/science.abd4385}. However, quantum statistics governing ultracold atom-molecule superradiance remain largely unexplored. Figure~\ref{phase2}(a) summarizes the quantum phases of ground-state molecules in $N$$\tilde{\Omega}$ parameter plane for $\hbar\tilde{\triangle}_{c}/E_{L}=4\times10^{3}$. Remarkably, the total atom number plays a crucial role in atom-molecule superradiant QPT. As expected, the larger values of $N$ correspond to smaller threshold with $\tilde{\Omega}_{\rm cr} \sim N^{-1/2}$. The slight deviation between analytic threshold (red dashed line) and numerical result (blue solid line) for large $N$ is ascribe to PA field induced strong spatially dependent two-body interaction of atoms. We emphasize that the superradiance is robust against small variations in short-range collisions. Over a broad range of $a_{mm}$, both $|\alpha|$ and $N_m$ gradually decrease as the $s$-wave scattering length increases, as shown in Fig.~\ref{phase2}(b).

To gain deeper insight into atom-molecule superradiant, we plot the cavity field amplitude $|\alpha|$ as function of total atom number $N$ for $\hbar\tilde{\Omega}/E_L=2$ [Fig.~\ref{phase2}(c)]. Clearly, the net cavity amplitude is proportional to $N^{3/2}$, which yields steady-state photon  intensity $N_s\sim N^3$. This cubic scaling of superradiance for hybrid quantum systems, originating from bosonic enhancement, is fundamentally different from the extensively studied ultracold bosonic  and fermionic atomic superradiance due to distinct quantum statistics~\cite{Ritsch_2021,Baumann2010,doi:10.1126/science.abd4385}. Remarkably, this distinct scaling behavior offers a new method for diagnosing molecular states via quantum nondemolition measurements. Furthermore, the strong sensitivity of $N_s$ to $N$ enhances our understanding of quantum superchemistry and provides new insights into controlling many-body chemical reactions~\cite{doi:10.1126/science.aam6299,doi:10.1126/science.abl7257,doi:10.1126/science.abn8525}.

 {\em Photon-matter entanglement}.--- In above discussion, the cavity-matter coupling is in weak-coupling regime with $\tilde{\Omega}/\kappa <0.1$ and $\tilde{\Omega}/\tilde{\Delta}_{c}\ll 1$, allowing the treatment of cavity and matter-wave fields as coherent states. Nevertheless, strong photon-matter entanglement can still emerge from tripartite interaction, akin to generation of entangled states, such as twin-Fock states via weak spin-exchange collisions in cold atoms~\cite{doi:10.1126/science.aag1106}. Neglecting system dissipation, the hybrid system conserves two quantities satisfying the commutation relations, $[\hat{N}, \hat {\cal H}_2]=0$ and $[\hat{N}_e, \hat {\cal H}_2]=0$. The ground state is expressed in terms of photon number $n$, $|\psi\rangle = \sum_n c_n|2(N_e-n)-N,N+n-N_e,n\rangle$. Here $2(N_e-n)-N$, $N+n-N_e$, and $n$ present the numbers of particles in atoms, molecules, and photons, respectively. The photon number is constrained by $\max(0,N_e-N)\leq N_a \leq N_e-N/2$ ensuring $N_b,N_m,N_a\geq0$. This representation significantly reduces Hilbert space dimension required to calculate ground states. 
 
To simplify analysis, we treat the system as a bipartite entity, decomposed it into light (A) and matter (M) subsystems. Photon-matter entanglement is quantified using the second R\'{e}nyi entropy by tracing out the photon degrees of freedom, $S_2=-\log\text{Tr}(\rho_{\rm M}^2)=-\log\sum_n |c_n|^4$ with $\rho_{\rm M}=\text{Tr}_A(|\psi\rangle\langle\psi|)$ being the reduced density matrix. In Fig.~\ref{S2NaNm}(a), we map the second R\'{e}nyi entropy $S_2$ on the $\tilde \Delta_c$-$\tilde \Omega$ parameter plane for fixing $N=10^4$. Remarkably, $S_2$ increases from zero to a finite value as $\tilde \Omega$ exceeds the critical Raman coupling. The phase boundary with $S_2 \approx 0$ aligns closely with the analytical result (red dashed line). As for SQL phase, photon-matter entanglement grows rapidly as $\tilde \Omega$ increases, providing a clear signature of superradiant QPT. 

The phase diagram for photon and molecule excitations mirror that of $S_2$. To illustrate this, we plot photon number $N_s$ and molecule number $N_m$ as functions of $\hbar\tilde\Omega/E_L$ in Fig.~\ref{S2NaNm}(b).The steady-state solutions from the mean-field (MF) approach agree excellently with exact diagonalization (ED) results, validating the validity of ground-state structures derived from the mean-field Gross-Pitaevskii equations. This result is further supported by examining quantum statistics of system. For SQL phase, the second-order autocorrelation function $g_{aa}^{(2)}(0)$ for cavity and crosscorrelation function $g_{am}^{(2)}(0)$ between cavity and molecule fields both equal 1,confirming that these field behave as coherent states. However, the light-matter entanglement becomes increasingly significant in the superradiance phase.
 
Remarkably, we find that $g_{aa}^{(2)}(0)=2$ for N phase, indicating thermal photon statistics. This counterintuitive result can be understood from the tripartite Hamiltonian $\hat {\cal H}_2$. Under the undepleted pump approximation, where the atomic field is treated as a classical source by replacing $\hat{b}  \rightarrow \sqrt{N}$, the reduced parametric conversion process will generate entangled molecule-photon pairs by leveraging atom-molecule superradiance. When $\tilde{\Omega}<\tilde{\Omega}_{\rm cr}$, N phase is characterized by the two-mode squeezed vacuum state 
\begin{align}
|\psi_S \rangle &= \frac{1}{\cosh r} \sum_{n=0}^{\infty} (-\tanh r)^n |n, n \rangle,
\end{align}
where the squeezing parameter $r$ satisfies $\tanh r=[\tilde{\Delta}_{c}+\delta'-\sqrt{(\tilde{\Delta}_{c}+\delta')^2-N\tilde{\Omega}^2}]/\sqrt{N\tilde{\Omega}^2}$. This state represents a coherent superposition of strictly particle number correlated Fock states. Clearly, both cavity and molecule modes exhibit thermal quantum statistics with $g_{aa}^{(2)}(0)=g_{mm}^{(2)}(0)=2$, corresponding to the ground state excitations $N_s=N_m= \tanh^2r/(1- \tanh^2r)$. Interestingly, the photon-molecule pair is strongly correlated with $g_{am}^{(2)}(0)=  2+1/\sinh^2r \gg 1$. This implies that the generated two-mode squeezing between photon and long-lived ultracold molecule offers potential applications in quantum metrology, particularly under decoherence.

{\em Conclusion}.---We have proposed an experimental scheme to explore atom-molecule superradiance using cavity-enhanced two-photon PA,  enabling the realization of strong long-range three-body interactions in hybrid matter-wave fields of atoms and molecules. The ground-state structures of cavity coupled hybrid atom-molecule condensate were systematically investigated. It is shown that the self-organized square lattice phase, governed by a novel tripartite many-body interaction. This phase is associated with second-order superradiant QPT featuring spontaneous $U(1)$ symmetry breaking. Distinct from ultracold atom superradiance, we observe $N^3$ scaling of photon number with total atom number, highlighting bosonic enhancement with distinct quantum statistics. This result provides an unambiguous smoking gun of generating molecules from ensembles of ultracold atoms. Furthermore, the strong photon-matter entanglement generated in atom-molecule superradiance may facilitate the study of entanglement-enhanced metrology~\cite{doi:10.1126/science.aag1106}. 

{\em Acknowledgments}.---This work was supported by the NSFC (Grants No. 12274473 and No. 12135018), by the National Key Research and Development Program of China (Grant No. 2021YFA0718304), by the Strategic Priority Research Program of CAS (Grant No. XDB28000000).



\newpage{}
\begin{widetext}
\begin{center}
\textbf{\large Supplementary materials: Atom-Molecule Superradiance and Entanglement with Cavity-Mediated Three-Body Interactions
}
\end{center}
\tableofcontents

 \setcounter{equation}{0} \setcounter{figure}{0} \setcounter{table}{0} %
 \renewcommand{\theequation}{S\arabic{equation}} \renewcommand{\thefigure}{S%
  \arabic{figure}} \renewcommand{\bibnumfmt}[1]{[S#1]}

~
~
~


\section{The cavity mediated atom-molecule Hamiltonian}\label{appA}

In this section, we present the detailed derivation of the cavity mediated atom-molecule Hamiltonian for the experimental setup schematic and level diagram displayed in Fig. 1 of the main text.
Under the rotating-wave approximation, the Hamiltonian except the kinetic energy and the two-body s-wave collisional interaction is given by
\begin{eqnarray}\label{h1}
\hat{h}_{1}/\hbar=&&\Delta_{c}\hat{a}^{\dagger}\hat{a}+\delta\hat{m}^{\dagger}\hat{m}+\Delta\hat{e}^{\dagger}\hat{e}+[\Omega(\mathbf{r})\hat{b}^{2}\hat{e}^{\dagger}+g^{*}(\mathbf{r})\hat{a}^{\dagger}\hat{m}^{\dagger}\hat{e}+\rm {H.c.}],
\end{eqnarray}
where $\hat{a}$ is the annihilation operator for the optical cavity photon, and $\hat{b}$ and $\hat{m}$ ($\hat{e}$) are annihilation operators for the atom and the ground state (quasi-bound) molecule, respectively.
Meanwhile, $\Delta_{c}$ is the pump-cavity detuning, and $\Delta$ ($\delta$) is the sigle (two)-photon detuning.
In addition, $\Omega(\mathbf{r})=\Omega_{0}\cos(k_{L}y)$ and $g(\mathbf{r})=g_{0}\cos(k_{L}x)$ are the spatially dependent coupling strength of laser field and cavity field, respectively.

Then the Heisenberg equations of motion for the atom, molecule and cavity operators read
\begin{eqnarray}
i\dot{\hat{a}}=&&	\Delta_{c}\hat{a}+g_{0}\cos(k_{L}x)\hat{m}^{\dagger}\hat{e},\nonumber\\
i\dot{\hat{b}}=&&2\Omega_{0}\cos(k_{L}y)\hat{b}^{\dagger}\hat{e},\nonumber\\
i\dot{\hat{m}}=&&\delta\hat{m}+g_{0}\cos(k_{L}x)\hat{a}^{\dagger}\hat{e},\nonumber\\
i\dot{\hat{e}}=&&\Delta\hat{e}+\Omega_{0}\cos(k_{L}y)\hat{b}^{2}+g_{0}\cos(k_{L}x)\hat{a}\hat{m}.
\end{eqnarray}
In the dispersive regime $\Delta\gg\{g_{0},\Omega_{0}\}$, the quasi-bound molecular state $|e\rangle$ can be adiabatically eliminated since its dynamics reaches quickly to a steady state with a negligible population, which leads to
\begin{eqnarray}
\hat{e}=-\frac{\Omega(\mathbf{r})\hat{b}^{2}+g(\mathbf{r})\hat{a}\hat{m}}{\Delta}.
\end{eqnarray}
Consequently, the Hamiltonian in Eq.~\eqref{h1} becomes
\begin{eqnarray}
\hat{h}_{2}/\hbar=&&\Delta_{c}\hat{a}^{\dagger}\hat{a}+[\delta+U_{0}\cos^{2}(k_{L}x)\hat{a}^{\dagger}\hat{a}]\hat{m}^{\dagger}\hat{m}+U_{y}\cos^{2}(k_{L}y)\hat{b}^{\dagger 2}\hat{b}^{2}+\Omega\cos(k_{L}x)\cos(k_{L}y)(\hat{a}\hat{b}^{\dagger 2}\hat{m}+\hat{a}^{\dagger}\hat{b}^{2}\hat{m}^{\dagger}),
\end{eqnarray}
where $\Omega=-\frac{g_{0}\Omega_{0}}{\Delta}$ is the two-photon scattering between pump and cavity mode and $U_{0}=-\frac{g_{0}^{2}}{\Delta}$ ($U_{y}=-\frac{\Omega_{0}^{2}}{\Delta}$) is the optical Stark shift of the cavity (pump) field.
Taking into account the kinetic energy and the two-body s-wave collisional interaction, the many-body interaction Hamiltonian for atom-molecule cavity system is given by
\begin{eqnarray}\label{manybodyh}
\hat {\cal H}_0&=
\hbar\Delta_c\hat{a}^{\dag}\hat{a} + \sum_{\sigma}\int d{\bf{r}}\hat{\psi}_{\sigma}^{\dagger}(\mathbf{r})[\frac{{\bf p}^{2}}{2m_{\sigma}}+V_e+U_{\sigma}]\hat{\psi}_{\sigma}(\mathbf{{r}}),\nonumber\\
&+\frac{1}{2}\sum_{\sigma\sigma'}\int d{\bf{r}} (g_{\sigma\sigma'}+\lambda_{\sigma\sigma{'}})\hat{\psi}_{\sigma}^{\dagger}(\mathbf{r})\hat{\psi}_{\sigma'}^{\dagger}(\mathbf{{r}})\hat{\psi}_{\sigma'}(\mathbf{{r}})\hat{\psi}_{\sigma}(\mathbf{{r}}),\nonumber\\
&+\hbar\Omega\int d{\mathbf{r}} \cos(k_{L}x)\cos(k_{L}y)\hat{a}\hat{\psi}_{b}^{\dagger2}(\mathbf{r})\hat{\psi}_{m}(\mathbf{r}) 
+{\rm H.c.},\label{manyh} 
\end{eqnarray}
where $\hat{\psi}_{b}$ and $\hat{\psi}_{m}$ are the annihilation operators for the atomic and molecular fields with the masses satisfying $m_m=2m_b$. 
And the term $U_{m}=\hbar[\delta+U_{0}\cos^{2}(k_{L}x)\hat{a}^{\dagger}\hat{a}]$ contributes as an effective trap potential for molecules. Meanwhile, the PA field induces an effective two-body interaction for atoms, $\lambda_{bb}= 2\hbar U_y \cos^{2}(k_{L}y)$, which can be used to fine tuning the collisional interaction avoiding the atom losses.
$V_e=m_{\sigma}\omega_{\perp}^2(x^2+y^2+\gamma^2z^2)/2$ is the external spin independent trapping potential with $\omega_{\perp}=(2\pi)130\rm{Hz}$ being the radial trap frequency and $\gamma=10$ being the trap aspect ratio. 
Moreover, the two-body short-range interaction $g_{{\sigma\sigma}}={4\pi\hbar^{2}a_{{\sigma\sigma}}}/{m_{\sigma}}$, $g_{\sigma\sigma'}={3\pi\hbar^{2}a_{bm}}/{m_{b}}$  with $a_{\sigma\sigma'}$ being $s$-wave scattering lengths for intraspecies ($\sigma=\sigma'$) and interspecies ($\sigma\neq\sigma'$) matter wave fields.
Then the dynamical equations for this system take the form
\begin{eqnarray}
i\dot{\hat{a}}=&&[\Delta_{c}+U_{0}\int d\mathbf{r}\cos^{2}(k_{L}x)\hat{\psi}_{m}^{\dagger}(\mathbf{r})\hat{\psi}_{m}(\mathbf{r})-i\kappa]\hat{a}+\Omega\int d\mathbf{r}\cos(k_{L}x)\cos(k_{L}y)\hat{\psi}_{b}^{2}(\mathbf{r})\hat{\psi}_{m}^{\dagger}(\mathbf{r}),\nonumber\\
i\dot{\hat{\psi}}_{m}=&&	[-\frac{\hbar\nabla^{2}}{4M_b}+\delta+U_{0}\cos^{2}(k_{L}x)\hat{a}^{\dagger}\hat{a}+\frac{1}{\hbar}V_{e}]\hat{\psi}_{m}+\Omega\cos(k_{L}x)\cos(k_{L}y)\hat{a}^{\dagger}\hat{\psi}_{b}^{2}(\mathbf{r})\nonumber\\&&+\frac{1}{\hbar}[g_{mm}\hat{\psi}_{m}^{\dagger}(\mathbf{r})\hat{\psi}_{m}^{2}(\mathbf{r})+g_{bm}\hat{\psi}_{b}^{\dagger}(\mathbf{r})\hat{\psi}_{b}(\mathbf{r})\hat{\psi}_{m}(\mathbf{r})],\nonumber\\
i\dot{\hat{\psi}}_{b}=&&	(-\frac{\hbar\nabla^{2}}{2M_b}+\frac{1}{\hbar}V_{e})\hat{\psi}_{b}(\mathbf{r})+2U_{y}\cos^{2}(k_{L}y)\hat{\psi}_{b}^{\dagger}(\mathbf{r})\hat{\psi}_{b}(\mathbf{r})^{2}+2\Omega\cos(k_{L}x)\cos(k_{L}y)\hat{a}\hat{\psi}_{b}^{\dagger}(\mathbf{r})\hat{\psi}_{m}(\mathbf{r})\nonumber\\&&+\frac{1}{\hbar}[g_{bb}\hat{\psi}_{b}^{\dagger}(\mathbf{r})\hat{\psi}_{b}^{2}(\mathbf{r})+g_{bm}\hat{\psi}_{m}^{\dagger}(\mathbf{r})\hat{\psi}_{m}(\mathbf{r})\hat{\psi}_{b}(\mathbf{r})].
\end{eqnarray}
In the far dispersive regime with $|\Delta_{c}/\kappa|\gg1$~\cite{RevModPhys.85.553, doi:10.1080/00018732.2021.1969727}, the cavity field quickly reaching a steady state is much faster than the external atomic motion, thus the cavity field can be adiabatical eliminated~\cite{Baumann2010, 10.1126/science.1220314}, which yields 
\begin{eqnarray}\label{cavity}
\hat{a}	=	-\frac{\Omega\int d\mathbf{{r}}\cos(k_{L}x)\cos(k_{L}y)\hat{\psi}_{b}^{2}(\mathbf{{r}})\hat{\psi}_{m}^{\dagger}(\mathbf{r})}{\tilde{\Delta}_{c}-i\kappa},	
\end{eqnarray}
where $\tilde{\Delta}_{c}=\Delta_{c}+U_{0}\int d\mathbf{r}\cos^{2}(k_{L}x)\hat{\psi}_{m}^{\dagger}(\mathbf{r})\hat{\psi}_{m}(\mathbf{{r}})$ is the effective dispersive shift of cavity .

Specially, an effective long-range three-body interaction is induced by integrating out the cavity field with the steady-state solution Eq.~\eqref{cavity}, which yields
\begin{eqnarray}
\hat{\cal H}_{1}/\hbar=\frac{\chi}{6}\int\int d\mathbf{r}d\mathbf{r}^{\prime}\mathcal{D}(\mathbf{r},\mathbf{r^{\prime}})\hat{\psi}_{m}^{\dagger}(\mathbf{r})\hat{\psi}_{b}^{\dagger 2}(\mathbf{r}^{\prime})\hat{\psi}_{m}(\mathbf{r}^{\prime})\hat{\psi}_{b}^{2}(\mathbf{r}),
\end{eqnarray}
where $\mathcal{D}(\mathbf{r},\mathbf{r^{\prime}})=\cos(k_{L}x)\cos(k_{L}x^{\prime})\cos(k_{L}y)\cos(k_{L}y^{\prime})$ is the long-range  potential and $\chi=-{12\tilde{\Delta}_{c}\Omega^{2}}/({\tilde{\Delta}_{c}^{2}+\kappa^{2}})$ is the tunable strength of three-body interactions.
Here, the effective three-body interaction is tunable by manipulating the external pump field and cavity field.

\begin{figure}
 \includegraphics[width=0.85\columnwidth]{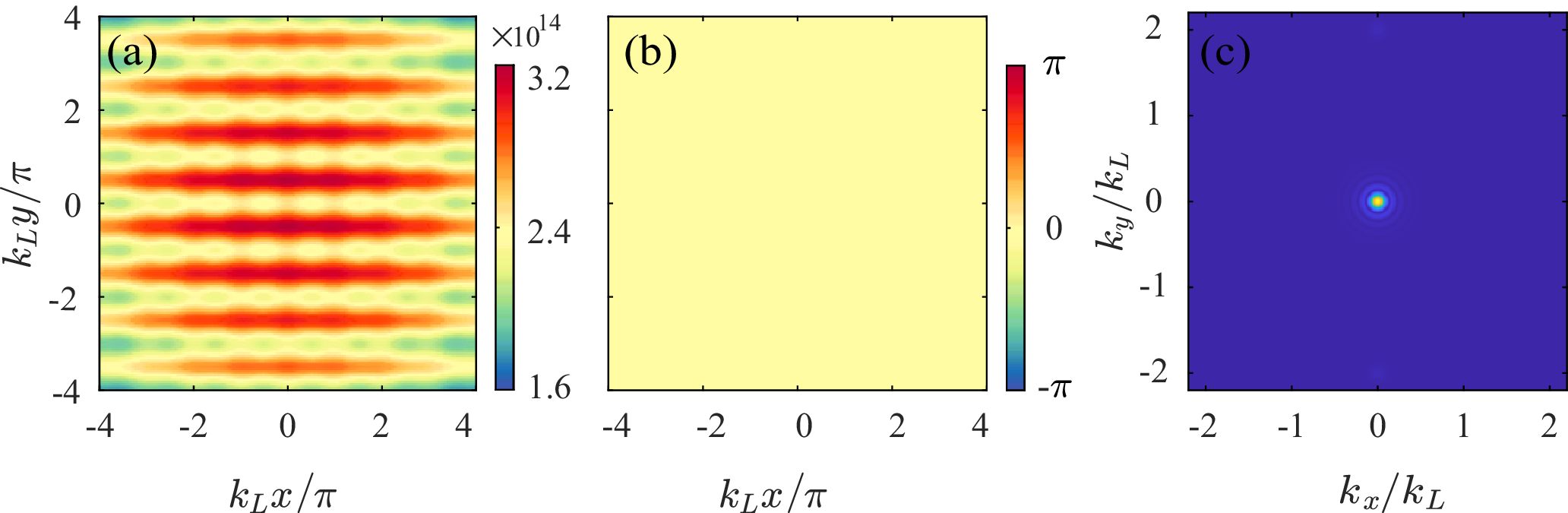} \caption{\label{SM_2}
(a) The density (units of $m^{-2}$) (b) phase and (c) momentum distribution of atoms. 
}
 \end{figure}
\section{Ground-states obtained via solving Gross-Pitaevskii equations}
In order to investigate the ground-state structures of atom-molecule cavity system, we employ the mean-field theory to self-consistently solve the Gross-Pitaevskii (GP) equations~\cite{PhysRevResearch.2.043318} with the aid of the steady-state solution of the intracavity amplitude.
Thus, the annihilation operators for atomic and molecular fields are substituted by the condensate wave functions $\psi_{\sigma}=\langle\hat{\psi}_{\sigma}\rangle$ and the steady-state photon amplitude $\alpha =\langle \hat{a} \rangle$ is self-consistently determined by the condensate wave functions. 
To utilize the commonly used imaginary time evolution, we can obtain the ground states of the condensate wave functions by numerically minimizing the free energy functional ${ \mathcal{E}}(\psi_m, \psi_b)=\langle \hat{\cal H}_0 \rangle$.
Specifically, we consider ultracold $^{133}$Cs atoms inside an optical cavity with dissipation rate $\kappa=50 E_L/\hbar$, where $E_L/\hbar=1.8~{\rm kHz}~(2\pi)$ is the single-photon recoil energy. 
Moreover, an quasi-two-dimensional harmonic-oscillator trapping potential $V_e=m_{\sigma}\omega_{\perp}^2(x^2+y^2+\gamma^2z^2)/2$ is implemented to confine the atomic condensate, where $\omega_{\perp}=(2\pi)130\rm{Hz}$ is the radial trap frequency and $\gamma=10$ is the trap aspect ratio.
The short-range two-body collisional interaction $g_{{\sigma\sigma}}={4\pi\hbar^{2}a_{{\sigma\sigma}}}/{m_{\sigma}}$ and $g_{\sigma\sigma'}={3\pi\hbar^{2}a_{bm}}/{m_{b}}$  with $a_{\sigma\sigma'}$ being $s$-wave scattering lengths for intraspecies ($\sigma=\sigma'$) and interspecies ($\sigma\neq\sigma'$) matter wave fields. In our simulation, we fix $a_{bb}=a_{bm}=50 a_B$ and $a_{mm}=100 a_B$ with $a_B$ being the Bohr radius.

Consequently, we identify a self-ordered superradiant phase for ground states of molecular condensate via numerically solving GP equations. As the atom-molecule conversion strength increases, a superradiant quantum phase transition (QPT) occurs. During the transition, the ground states of the atom-molecule cavity system shift from a normal phase ($\alpha=0$) to a superradiant phase ($\alpha\neq0$).
For this intriguing superradiant phase, the density profile of molecular wave function manifests a remarkable $\lambda/2$-period density modulation along both $x$ and $y$ axes. Owing to this characteristic pattern, we refer to this phase as square lattice (SQL) phase, as shown in Fig 2 (e) of the main text. 
And the phase profile of molecular wave function displays a staggered $\lambda$-period phase modulation, as illustrated in Fig 2 (f) of the main text. Notably, there exists a relative phase difference of $\pi$ between neighboring sites, which is associated with $U (1)$ symmetry breaking, and we will further explore in subsequent section. Unlike the molecular wave function, the density distribution of the atomic condensate wave function is structureless along the cavity axis, as displayed in Fig.~\ref{SM_2} (a). 
Meanwhile, the phase distribution of the atomic condensate wave function is also trivial, with a uniform distribution of zero shown in Fig.~\ref{SM_2} (b).
Moreover, the momentum-space distribution of atomic condensate wave function is given in Fig.~\ref{SM_2} (c), which reveals that a great amount of atoms are in zero-momentum states.
Notably, figuring out the ground-state properties of atomic and molecular condensate wave functions can facilitate a more profound exploration of the underlying physical insights.

\section{Effective potential and Goldstone mode for superradiant phase}
In this section, we give the detailed derivation of the effective potential of superradiant phase, and calculate the excitation spectrum to find the gapless Goldstone mode, which demonstrates the rigidity of the superradiant phase.
To gain important physical insight, we expand field operators by $\hat{\psi}_{b}=\sqrt{{1}/{V}}\hat{b}$ and $\hat{\psi}_{m}=2\sqrt{{1}/{V}}\cos(k_{L}x)\cos(k_{L}y)\hat{m}$ with $V$ being the volume of condensate in the single recoil scattering limit~\cite{Baumann2010, 10.1126/science.1220314, PhysRevLett.123.160404, PhysRevLett.121.163601}. This leads the many-body Hamiltonian, excluding the external trapping potential and the two-body interaction, to an Hamiltonian given by
\begin{eqnarray}\label{s9}
\hat{\cal H}_{2}/\hbar=\tilde{\Delta}_{c}\hat{a}^{\dagger}\hat{a}+\delta^{\prime}m^{\dagger}\hat{m}+{\frac{\tilde{\Omega}}{2{\sqrt{N}}}}(\hat{b}^{\dagger 2}{}\hat{a}\hat{m}+\rm H.c.),
\end{eqnarray}
where $N$ is the total atom number, and 
$\delta^{\prime}=\delta+E_{L}/\hbar$ with $E_{L}$ being the single-photon
recoil energy. Meanwhile, $\tilde{\Omega}={\xi\sqrt{\bar{n}_{2D}}\Omega}$, where $\bar{n}_{2D}$ is practically the 2D mean number density of atoms and $\xi=\int\phi_{bz}^{* 2}\phi_{mz}dz=( {\gamma}/{\pi})^{3/4}\sqrt{{2\pi\ell_{m}}/{\gamma(\ell_{b}^{2}+2\ell_{m}^{2})}}$
with $\phi_{\sigma z}(z)=(\gamma/\pi\ell_{\sigma}^{2})^{1/4}e^{-\gamma z^{2}/2\ell_{\sigma}^{2}}$ and $\ell_{\sigma}=\sqrt{{\hbar}/({M_{\sigma}\omega_{\perp}})}$.

Subsequently, we assume that $\hat{a}\rightarrow\alpha+\delta\hat{a}$, $\hat{m}\rightarrow\beta+\delta\hat{m}$ and $\hat{b}\rightarrow\sqrt{N-2m^{\dagger}\hat{m}}$, where $\delta\hat{a}$ and $\delta\hat{m}$ denote the photonic and molecular fluctuations of the system around its mean-field values with $\langle\hat{a}\rangle=\alpha$ and $\langle\hat{m}\rangle=\beta$ ~cite{PhysRevE.67.066203}
.
For simply, we adopt the notations $\delta\hat{a}\equiv\hat{a}$ and $\delta\hat{m}\equiv\hat{m}$, so $\hat{a}\rightarrow\alpha+\hat{a}$, $\hat{m}\rightarrow\beta+\hat{m}$ and $\hat{b}\rightarrow\sqrt{N-2|\beta|^{2}-2(\beta^{*}\hat{m}+\hat{m}^{\dagger}\beta+\hat{m}^{\dagger}\hat{m})}$.
As for the normal phase, it's obvious that $\alpha=\beta=0$.
To obtain the value of $\alpha$ and $\beta$ in ground state of superradiant phase, we submit above assumption into the Hamiltonian in Eq.~\eqref{s9}, and make coefficients of the linear terms be zero to minimum the ground-state energy, then we have
\begin{eqnarray}
\label{e6}
\alpha=&&{\frac{\tilde{\Omega}}{2\sqrt{N}\tilde{\Delta}_{c}}}(2|\beta|^{2}-N)\beta^{*},\nonumber\\
\label{e7}
|\beta|^{2}=&&\frac{N}{3}-\frac{1}{6}\sqrt{N^{2}+\frac{12N\delta^{\prime}\tilde{\Delta}_{c}}{\tilde{\Omega}^{2}}}.
\end{eqnarray}
%
Clearly, in the absence of cavity dissipation, the superradiant QPT occurs at a critical Raman coupling strength
\begin{eqnarray}
\tilde{\Omega}_{cr}	=2\sqrt{\frac{\delta^{\prime}\tilde{\Delta}_{c}}{N}}.
\end{eqnarray}

%

\begin{figure}
 \includegraphics[width=0.85\columnwidth]{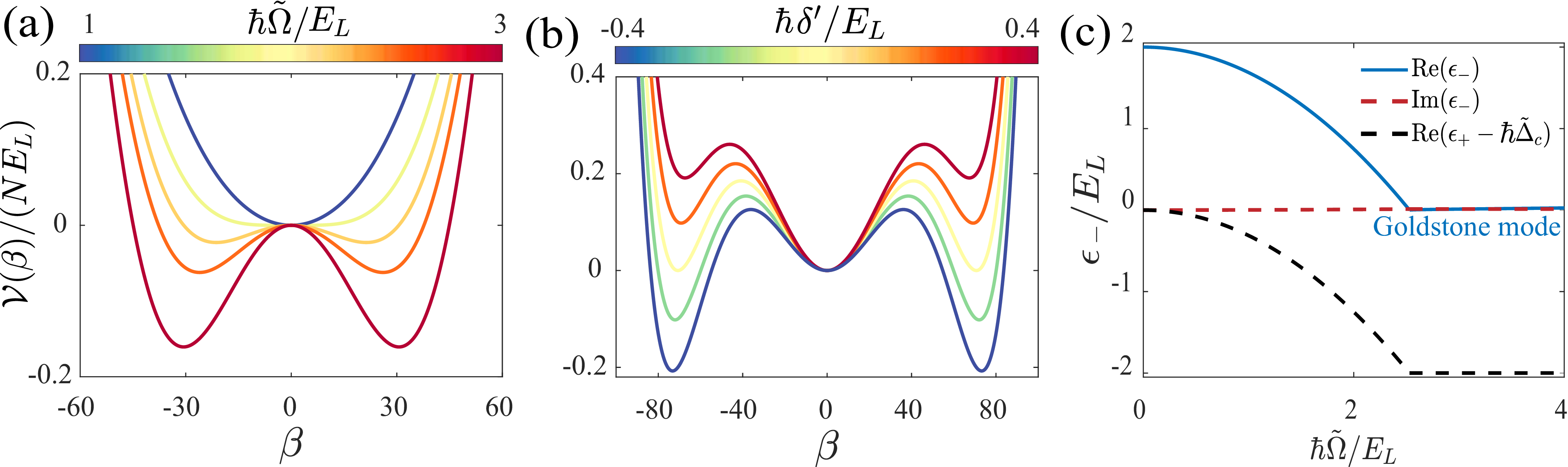} \caption{\label{potential_SM}
 (a) The effective potential ${\cal V}({\beta})$ as a function of $\beta$ for different $\hbar \tilde{\Omega}$ with parameters $\hbar \delta'/E_L=2$ and $\hbar \tilde{\Delta}_c/E_L=4\times 10^3$. (b) The effective potential ${\cal V}({\beta})$ as a function of $\beta$ for different $\delta'$ with parameters $\hbar \tilde{\Omega}/E_L=2$ and $\hbar \tilde{\Delta}_c/E_L=-4\times 10^3$. The total paticle number is $N=1\times 10^4$. (c) The lowest branch $\epsilon_{-}$ as a function of $\tilde{\Omega}$ with $\hbar \tilde{\Delta}_c/E_L=8\times 10^3$, $\hbar \delta'/E_L=2$ and nonzero cavity dissipation $\hbar \kappa= 50 E_L$.}
 \end{figure}
 
To further elucidate the mechanism of superradiant QPT, we derive the effective potential by integrating out the cavity field
\begin{eqnarray}\label{ev}
{\cal V}({\beta})/\hbar={\frac{\tilde{\Omega}^{2}}{N\tilde{\Delta}_{c}}}(N|\beta|^{4}-|\beta|^{6})+(\delta^{\prime}-{\frac{\tilde{\Omega}^{2}}{4\tilde{\Delta}_{c}}}N)|\beta|^{2},
\end{eqnarray}
 with $\beta=\langle\hat{m}\rangle$. For fixed phase ${\rm arg(\beta)=0}$, the effective potential ${\cal V}({\beta})$ at $\hbar \delta'/E_L=2$ and $\hbar \tilde{\Delta}_c/E_L=4\times 10^3$ as a function of order parameter $\beta$ is displayed in Fig.~\ref{potential_SM} (a) for a range of couplings $\tilde{\Omega}$. As $\tilde{\Omega}$ increases, the single minimum of ${\cal V}({\beta})$ at $\beta=0$ bifurcates into two symmetric local minima at $\beta \neq 0$, signaling the onset of a second-order QPT. Notably, ${\cal V}({\beta})$ possesses the $U(1)$ symmetry under the gauge transformation $\beta \rightarrow \beta e^{i\theta}$. When $\tilde{\Omega}>\tilde{\Omega}_{\rm cr}$, ${\cal V}({\beta})$ transitions from a single minimum at the origin to a sombrero shape potential with a circular valley of degenerate minima, as depicted in Fig. 1 (b) of the main tex. This effective potential can be directly measured using the condition $\alpha={\tilde{\Omega}}(2|\beta|^{2}-N)\beta^{*}/(2\sqrt{N}\tilde{\Delta}_{c})$.

In comparison to the atom-cavity superradiance characterized by the generalized Dicke mode, the additional term proportional to $|\beta|^6$
is emerged for hybrid atom-molecule system, which may exhibit a first-order QPT for negative pump-cavity detuning.
Analogously, the effective potential ${\cal V}({\beta})$ at $\hbar \tilde{\Omega}/E_L=2$ and $\hbar \tilde{\Delta}_c/E_L=-4\times 10^3$ as a function of order parameter $\beta$ is displayed in Fig.~\ref{potential_SM} (b) for a range of couplings $\delta'$.
As the value of $\delta'$ varies from negative to positive, the number of minima of effective potential initially changes from a single minimum to three minima and then further alters to two minima. Concurrently, the corresponding signs of these minima also shift from negative to positive. This characteristic behavior, where the minima's number and sign vary in such a coordinated manner, is a hallmark indication of a first-order QPT. Moreover, when $\delta'=0$, the effective potential presents three minima with zero values, signifying the first-order transition boundary. 

Furthermore, we calculate the collective excitations of cavity-coupled atom-molecule system. 
By employing the solutions of Eq.~\eqref{e6} and Eq.~\eqref{e7}, it becomes straightforward to obtain $\alpha={\tilde{\Omega}}{(2\ensuremath{\mu}-N)}\sqrt{\mu}e^{-i\theta}/(2\sqrt{N}\tilde{\Delta}_{c})$ and $\beta= \sqrt{\mu}e^{i\theta}$, where $\mu={N}/{3}-{\sqrt{N^{2}+{12N\delta^{\prime}\tilde{\Delta}_{c}}/{\tilde{\Omega}^{2}}}}/{6}$ is introduced to make expressions more clean.
Then the quadratic Hamiltonian describing collective excitations of superradiant phase takes the form
\begin{eqnarray}\label{s14}
\hat{\cal H}^{(2)}/\hbar=&&(\tilde{\Delta}_{c}-i\kappa)\hat{a}^{\dagger}\hat{a}+[\delta^{\prime}-2\frac{\tilde{\Omega}^{2}}{N\tilde{\Delta}_{c}}\ensuremath{(2\mu^{2}-\mu N)}]\hat{m}^{\dagger}\hat{m}+[{\frac{1}{2}}\frac{\tilde{\Omega}}{\sqrt{N}}(N-4\mu)\hat{a}\hat{m}
\nonumber\\&&-{\frac{\tilde{\Omega}^{2}}{2N\tilde{\Delta}_{c}}\ensuremath{(2\mu^{2}-\mu N)}}e^{-2i\theta}\hat{m}\hat{m}-\mu\frac{\tilde{\Omega}}{\sqrt{N}}e^{2i\theta}\hat{a}\hat{m}^{\dagger}+\rm H.c.],
\end{eqnarray}
where the cavity decay $\kappa$ is taken into account.
After the gauge transformations $\hat{a}\rightarrow\hat{a}e^{-i\theta}$ and $\hat{m}\rightarrow\hat{m}e^{i\theta}$, the Bogoliubov Hamiltonian reads
\begin{eqnarray}
\hat{\cal H}^{(2)}/\hbar=
&&\omega_{1}\hat{a}^{\dagger}\hat{a}+\omega_{2}\hat{m}^{\dagger}\hat{m}+(\Omega_{1}\hat{a}\hat{m}+\Omega_{2}\hat{m}\hat{m}+\Omega_{3}\hat{a}\hat{m}^{\dagger}+\rm H.c.),
\end{eqnarray}
where
\begin{eqnarray}
\omega_{1}	=&&	\tilde{\Delta}_{c}-i\kappa,\nonumber\\
\omega_{2}	=&&	\delta^{\prime}-2\frac{\tilde{\Omega}^{2}}{N\tilde{\Delta}_{c}}\ensuremath{(2\mu^{2}-\mu N)},\nonumber\\
\Omega_{1}	=&&	{\ensuremath{\frac{1}{2}}}\frac{\tilde{\Omega}}{\sqrt{N}}(N-4\mu),\nonumber\\
\Omega_{2}	=&&	-{\ensuremath{\frac{\tilde{\Omega}^{2}}{2N\tilde{\Delta}_{c}}}\ensuremath{(2\ensuremath{\mu^{2}}-\ensuremath{\mu}N)}},\nonumber\\
\Omega_{3}	=&&	-\mu\frac{\tilde{\Omega}}{\sqrt{N}}
\end{eqnarray}
are introduced for shorthand notation. As a result, the Heisenberg equations of motion for the quantum fluctuations in the photonic and atomic field operators can be obtained
\begin{eqnarray}
i\dot{\hat{a}}	=&&	\omega_{1}\hat{a}+\Omega_{1}\hat{m}^{\dagger}+\Omega_{3}\hat{m},\nonumber\\
i\dot{\hat{a}}^{\dagger}	=&&	-\omega_{1}^{*}\hat{a}^{\dagger}-\Omega_{1}\hat{m}-\Omega_{3}\hat{m}^{\dagger},\nonumber\\
i\dot{\hat{m}}	=&&	\omega_{2}\hat{m}+\Omega_{1}\hat{a}^{\dagger}+2\Omega_{2}\hat{m}^{\dagger}+\Omega_{3}\hat{a},\nonumber\\
i\dot{\hat{m}}^{\dagger}	=&&	-\omega_{2}\hat{m}^{\dagger}-\Omega_{1}\hat{a}-2\Omega_{2}\hat{m}-\Omega_{3}\hat{a}^{\dagger}.
\end{eqnarray}	
Furthermore, we recast these equations in the form of the Hopfield-Bogoliubov matrix
\begin{eqnarray}
\left(\begin{array}{cccc}
\omega_{1} & \Omega_{3} & 0 & \Omega_{1}\\
\Omega_{3} & \omega_{2} & \Omega_{1} & 2\Omega_{2}\\
0 & -\Omega_{1} & -\omega_{1}^{*} & -\Omega_{3}\\
-\Omega_{1} & -2\Omega_{2} & -\Omega_{3} & -\omega_{2}
\end{array}\right)\left(\begin{array}{c}
\hat{a}\\
\hat{m}\\
\hat{a}^{\dagger}\\
\hat{m}^{\dagger}
\end{array}\right)	=\epsilon	\left(\begin{array}{c}
\hat{a}\\
\hat{m}\\
\hat{a}^{\dagger}\\
\hat{m}^{\dagger}
\end{array}\right).
\end{eqnarray}
Thus the collective excitation spectra of our system can be conveniently calculated by numerically diagonalizing the Hopfield-Bogoliubov matrix $\hat{H}^{(2)}$.

As for the normal phase with $\alpha=\beta=0$, the quadratic Hamiltonian is given by
\begin{eqnarray}
\hat{\cal H}_{N}^{(2)}/\hbar=&&	\omega_{1}\hat{a}^{\dagger}\hat{a}+\delta^{\prime}\hat{m}^{\dagger}\hat{m}+\chi\hat{a}\hat{m}+\chi\hat{a}^{\dagger}\hat{m}^{\dagger}
\end{eqnarray}
with ${\ensuremath{\chi}=\ensuremath{\frac{1}{2}}}\sqrt{N}\tilde{\Omega}$.
Then the Heisenberg equations of motion of the quantum fluctuations in the photonic and atomic field operators are given by
\begin{eqnarray}
i\dot{\hat{a}}	=&&	\omega_{1}\hat{a}+\chi\hat{m}^{\dagger},\nonumber\\
i\dot{\hat{a}}^{\dagger}	=&&	-\omega_{1}^{*}\hat{a}^{\dagger}-\chi\hat{m},\nonumber\\
i\dot{\hat{m}}	=&&	\delta^{\prime}\hat{m}+\chi\hat{a}^{\dagger},\nonumber\\
i\dot{\hat{m}}^{\dagger}	=&&	-\delta^{\prime}\hat{m}^{\dagger}-\chi\hat{a}.
\end{eqnarray}
Analogously, we can recast these equations in the form of the Hopfield-Bogoliubov matrix
\begin{eqnarray}
\left(\begin{array}{cccc}
\omega_{1} & 0 & 0 & \chi\\
0 & \delta^{\prime} & \chi & 0\\
0 & -\chi & -\omega_{1}^{*} & 0\\
-\chi & 0 & 0 & -\delta^{\prime}
\end{array}\right)\left(\begin{array}{c}
\hat{a}\\
\hat{m}\\
\hat{a}^{\dagger}\\
\hat{m}^{\dagger}
\end{array}\right)	=	\epsilon\left(\begin{array}{c}
\hat{a}\\
\hat{m}\\
\hat{a}^{\dagger}\\
\hat{m}^{\dagger}
\end{array}\right).
\end{eqnarray}
Then we calculate its eigenvalues
\begin{eqnarray}
\epsilon_{-}^{\prime}	=&&	(\omega_{1}-\delta^{\prime}-\sqrt{(\delta^{\prime}+\omega_{1})^{2}-4\chi^{2}})/2,\nonumber\\
\epsilon_{+}	=&&	(\omega_{1}-\delta^{\prime}+\sqrt{(\delta^{\prime}+\omega_{1})^{2}-4\chi^{2}})/2,\nonumber\\
\epsilon_{+}^{\prime}	=&&	(\delta^{\prime}-\omega_{1}^{*}-\sqrt{(\delta^{\prime}+\omega_{1}^{*})^{2}-4\chi^{2}})/2,\nonumber\\
\epsilon_{-}	=&&	(\delta^{\prime}-\omega_{1}^{*}+\sqrt{(\delta^{\prime}+\omega_{1}^{*})^{2}-4\chi^{2})})/2.
\end{eqnarray}
It is noteworthy that the collective excitations of the system always have two positive and two negative eigenvalues due to the commutation relations of the creation and annihilation operators~\cite{PhysRevLett.112.173601}.
Among them, $\epsilon_{+}$ and $\epsilon_{-}$ are two positive eigenvalues, and $\epsilon_{+}$ ($\epsilon_{-}$) denotes the higher (lower) branch of collective excitation.

Furthermore, the characteristic collective excitations of the atom - molecule cavity system, presented as a function of the Raman coupling  $\tilde{\Omega}$, are illustrated in [Fig.~\ref{potential_SM} (c)].
We find that the energy gap between the higher and lower branches satisfies $\epsilon_{+}-\epsilon_{-}\approx \hbar \tilde{\Delta}_c \gg \hbar \omega_2$, which implies that the higher branch $\epsilon_{+}$ is effectively decoupled from the ground state of the lower branch $\epsilon_{-}$.
It is evident from our analysis that, within our model, the low-energy excitation hosts a gapless Goldstone mode when the Raman coupling strength $\tilde{\Omega}$ exceeds the threshold of the superradiant QPT.
This Goldstone mode emerges as a consequence of the spontaneously broken continuous $U (1)$ symmetry. Additionally, we have verified that this zero-energy mode remains approximately undamped, even in the presence of nonzero cavity dissipation, given that $\kappa/\tilde{\Delta}_c \ll 1$. Such findings contribute to a deeper understanding of the dynamical properties and symmetries within the atom-molecule cavity system.

\end{widetext}

\newpage

%

\end{document}